\documentclass[submission]{eptcs}

\usepackage[utf8]{inputenc}
\usepackage{graphicx}
\usepackage{booktabs}
\usepackage{footnote}
\usepackage{textcomp}
\usepackage{breakurl}
\usepackage{todonotes}
\usepackage{enumitem}
\usepackage{rotating}

\title{How to Put Usability into Focus:\\ Using Focus Groups to
  Evaluate the Usability of\\ Interactive Theorem Provers\thanks{The
    work presented here is part of the project Usability of Software
    Verification Systems within the BMBF-funded programme Software
    Campus.}}
\author{Bernhard Beckert \email{beckert@kit.edu} \and 
  Sarah Grebing \email{sarah.grebing@kit.edu} \institute{Karlsruhe Institute of Technology (KIT)} \and 
  Florian B\"{o}hl \email{boehl@kit.edu}}
\def\authorrunning{B. Beckert, S. Grebing \& F. B\"{o}hl}
\def\titlerunning{How to Put Usability into Focus: Using Focus Groups
  to Evaluate the Usability of ITPs}

\newcommand{\KeY}{Ke\kern-0.1emY}

\begin{document}

\maketitle

\begin{abstract}
In recent years the effectiveness of interactive theorem provers has 
increased to an extent that the bottleneck in the interactive process shifted 
to efficiency: while in principle large and complex theorems are provable 
(effectiveness), it takes a lot of 
effort for the user interacting with the system (lack of efficiency). 
We conducted focus groups to evaluate the usability of
Isabelle/HOL and the \KeY{} system with two goals: (a)~detect usability issues 
in the interaction between interactive theorem provers and their user, 
and (b)~analyze how evaluation and survey methods commonly used in the area of 
human-computer interaction, such as focus groups and 
co-operative evaluation, are applicable to the specific field of interactive 
theorem proving~(ITP). 

In this paper, we report on our experience using the evaluation method
focus groups and how we adapted this method to ITP. We describe our
results and conclusions mainly on the ``meta-level,'' i.e., we focus
on the impact that specific characteristics of ITPs have on the setup
and the results of focus groups. On the concrete level, we briefly summarise 
insights into the usability of the ITPs used in our case study.
\end{abstract}

\newcommand{\xparagraph}[1]{\textbf{#1}}
\newcommand{\setup}[0]{\emph{Our Setup:\ }}
\newcommand{\comment}[1]{\emph{Comment:\ #1}}

\section{Introduction}

In recent years, the effectiveness of interactive theorem provers~(ITPs) has 
increased to an extend that the bottleneck in the interactive process shifted to 
efficiency. 
While in principle large theorems are provable (effectiveness), it takes a lot 
of effort for the user interacting with the system (efficiency). 
This issue is recognized by the ITP-community and improvements are being 
developed. However, in contrast to properties like soundness or completeness, 
where rigorous methods are applied to provide evidence, the 
evidence for a better usability is lacking in many cases. 

The work reported here is part of a project where we apply various 
survey and evaluation 
methods
commonly used in the field of human-computer-interaction~(HCI) to ITPs,
including focus group discussions, usability testing, and user
experience questionnaires. Since expertise from both fields (ITP and
HCI) is required, we cooperate with user experience experts from
DATEV~eG within the BMBF-funded Software Campus programme (they are
well-versed in the ergonomic evaluation of standard software).

Our contribution in this paper is a description of how to use focus
group discussions, or \emph{focus groups} for short, to evaluate the
usability of ITPs. Moreover,  we explain what kind of questions one can expect
to answer using this focus groups, which are structured group
discussions guided by a moderator.

We report on two experiments we have
conducted, applying the 
focus group method to two different ITPs: the tactical theorem prover
Isabelle/HOL~\cite{Nipkow-Paulson-Wenzel:2002} and the interactive
program verification system \KeY{}~\cite{KeYBook2007}.  

We describe our setup for the focus group discussions, what needs
to be done in preparation for the discussions, and how the discussions
are evaluated to draw conclusions in the post-processing phase.
We hope that our experiences help others to conduct focus groups in this
field to improve the quality of ITP user interfaces.

\xparagraph{Related work.}

As mentioned above, the ITP community has noticed the need to evaluate
and improve usability, but so far structured usability evaluation
methods have rarely been applied to ITPs. Related work where this has
been done, includes the following:
In previous work~\cite{compare2012}, we have performed a
questionnaire-based evaluation of the \KeY{} system with a
questionnaire based on Green and Petre's cognitive dimensions
questionnaire~\cite{CDQuestionnaire} 
in order to get a first impression of the user's
perception and to develop first hypotheses about the usability of the
\KeY{} system.
Kadoda et~al.\ evaluated proof systems using Green and Petre's
Cognitive Dimensions questionnaire to develop a
list of desirable features for educational theorem
provers~\cite{kadodaDesirableFeatures}.
Aitken and Melham evaluated the interactive proof systems Isabelle and HOL 
using recordings of user interactions with the systems in collaboration with 
HCI experts. During the proof process the users were asked to think-aloud and 
after the recordings the users were interviewed.
Their goal was to study the activities performed by users of interactive 
provers during the proof process to obtain an interaction model of the users. 
As usability metric they propose to use typical user errors and 
they compared provers w.r.t.\ these errors~\cite{aitken, aitken95, 
aitken98}.
Based on the evaluation results also suggestions for improvements of the 
systems have been proposed by the authors including, besides others, improved 
search mechanisms and improved access to certain proof relevant components.
Jackson et~al.\ used co-operative evaluation methods on the CLAM Proof
Planner.  Users were asked to perform predefined tasks while using the
``think aloud technique'' to comment on what they where 
doing~\cite{ProofCritics}.
Questionnaires and interviews were used by Vujosevic and Eleftherakis to answer 
the question why
Formal Methods Tools are not used in industry~\cite{ImprovingFMTools}.
In the context of this work, also usability aspects of several formal
methods tools, such as the Alloy Analyzer, were evaluated.
For improving the interface of the prover NuPRL, a self-designed
questionnaire was used to evaluate the users' perceptions of the
interface~\cite{TPUsability}.

\xparagraph{Structure of this paper.}  We give an introduction to the
focus group method in Section~\ref{sec:focus-groups} and describe our
experiments using focus groups to evaluate the usability of ITPs. The
script used to guide the discussions is discussed in
Section~\ref{sec:script}. In Section~\ref{sec:results}, we present the
results of the experiments. Here, we focus on the meta-level, i.e., 
on the impact that specific characteristics of ITPs have on the setup
and the results of focus groups. On the concrete level, we briefly summarise 
insights into the usability of the ITPs used in our case study.
Finally, in Section~\ref{sec:conclusions}, we discuss
future work.

\section{Survey Method and Adaptation to ITPs} 
\label{sec:focus-groups}

\xparagraph{Focus Groups.} A \emph{focus group} is a discussion of five to ten
participants guided by a moderator. The moderator uses a prepared script to
initiate and structure the discussion which typically lasts about one to two
hours. Focus groups are a standard method in many areas to explore opinions
about specific products or topics, e.g., in market research. In the field of
human-computer interaction, they are used to explore user perspectives on
software systems and their usability.

\xparagraph{The Challenge of Conducting a Focus Group.}
Focus groups require less participants than evaluations using questionnaires and
the effort for conducting the discussion is less than that of one-on-one
interviews~\cite{UsabilityBook,ParticipatoryMethodsToolkit}. Still, it is a
non-trivial task to conduct a focus group. The discussions have to be
well-structured as well as lively and open to be productive. And it is a
challenge to steer the discussions towards the topics of interest without
predisposing possible answers or biasing the results in other ways. Fortunately,
we can draw from a big body of knowledge on how to conduct focus group
discussions (e.g.,~\cite{NielsenBook,Caplan1990}).  

\xparagraph{Focus Groups for ITPs.}
We have conducted two focus group discussions on the usability of ITPs, one for
the \KeY{} system~\cite{KeYBook2007} and one for
Isabelle/HOL~\cite{Nipkow-Paulson-Wenzel:2002}. The setup and topics of both
discussions were the same to make the results easier to compare. In the
remainder of this section we describe our course of actions specifically for
ITPs. We discuss the three phases of an evaluation using focus groups,
namely pre-processing, the discussion itself, and the post-processing.

\subsection{Pre-Processing}
\xparagraph{Selecting the Participants.}
In general, the composition of the focus group should be representative for the
user base of the tool being evaluated. But participants may also be selected
from certain sub-groups, such as beginners or experts. Both the level of
expertise in the relevant domain and the experience level for the evaluated tool
are relevant criteria. It is also crucial to have a group of participants who
are motivated and keen to debate.

\setup{}
The participants for our experiments were recruited using personal contacts to 
the relevant communities. We ensured that each group included novice,
intermediate, and expert users in different proportions. 
Besides that, the only criterion for
selection was that participants had to be open about the idea of focus
group discussions (mostly they were interested in a new experience and
to learn something new about using their tools). Most participants
were Master or PhD students, who had used \KeY\ resp.\ Isabelle for
their thesis work. We reimbursed participants' travel expenses but
they were not paid a fee. The \KeY\ group had seven and the Isabelle
group five participants. In the Isabelle group we had one novice, two 
intermediate and two experts users. In the \KeY{} group we had one novice, two 
intermediate and four expert users. 

\xparagraph{The Moderator.}
The moderator must not be one of the stake holders and must be neutral
in his or her opinion about the evaluated software. This excludes, for
example, developers of the evaluated tool. Nevertheless, the
moderator must understand the issues that are discussed. A well-prepared
and experienced moderator can greatly improve the results of a focus group
discussion.

\setup We had two different moderators, one 
for each discussion. Both were
computer scientists working in academia but not in the area of ITP. As
they were not expert moderators, they got an extensive training and
briefing prior to the discussions.

\xparagraph{The Technical Setup.}  It is advisable to use two adjacent
rooms, one for the discussion itself and one for observers, including
the experimenters and some domain experts (e.g., developers of the
ITP).  The discussion is recorded with at least one camera and several
microphones and is transmitted live to the observation room. This
setup has to be well tested beforehand as any technical problem can
seriously effect the post-processing of the recorded discussion.  It
is useful to provide a feedback channel from the observation room to
the moderator (using a headset) to give hints and provide relevant domain 
knowledge.

\setup We followed the advices about the spatial setup and 
used two adjacent rooms (one for the discussion, one for the
observers) with a glass window between them. The technical equipment
consisted of one camera and four microphones for recording, a back channel
from the observers to the moderator's headset, and lecture recording
software capable of recording and live streaming.

\xparagraph{The Script.}
The script contains all tasks and all questions for the focus group.
Only neutral questions can be asked explicitly (e.g., ``Please name
one good and one bad feature of the tool.''). Non-neutral questions
such as ``Is feature $X$ useful?'' are included in the script but are
not asked explicitly. Instead it is the moderator's task to guide the
discussion into the direction of these questions, e.g., by digging
deeper when a participant brings up a certain issue. Of course, the
moderator has to carefully balance neutrality and the desire to steer
the discussion in a certain direction.  The topics in the script
should build on each other in a meaningful way, e.g., from a general
topic to a specific topic~\cite{Caplan1990}. 

\setup The questions of the scripts for
our conducted focus groups are described in Sec.~\ref{sec:script}. The planned
duration for both groups was 2~hours. Due to lively
discussions, the actual duration was 2.5 resp.\ 3~hours.

\subsection{The Discussion}
The discussion itself starts with a round-robin introduction of the
participants and some small warm-up tasks, and it ends with a
cool-down task that allows to summarize the content of the discussion. The main
part consists of sub-discussions that are related to specific topics
such as usability aspects, tool features, etc. Each topic is
introduced by the moderator, possibly using example problems, mock-ups
of new features or similar material. After the recorded part of the
discussion ends, there should be time for questions and feedback from
the participants and the moderator (even if that part is not recorded, it is 
useful to take notes).

\setup Our discussion had three stages: the warm-up stage, the main stage and 
the cool-down stage.

The discussion was carried out according to the script, which is explained in 
detail in Sec.~\ref{sec:script}. All in all both discussion groups were lively 
and the participants engaged well in the discussion. Our impression was that 
the participants were open towards this method and were upfront about their 
systems. Our moderators were not experienced with moderations tasks and sometimes 
asked suggestive questions like ``Do you all share person A's opinion?''. While
this required an extra-careful analysis of the transcribed material, the 
damaging effect of such questions for the results was minimal. A thorough
analysis of the video material showed that often the group or certain
participants confirmed or denied a statement before the corresponding suggestive
question occurred. 

After the discussion, all participants had the chance to have a small offline 
talk with all project members and ask questions as well as express opinions 
about the focus group without being recorded.
We believe these offline discussions were a good opportunity for the 
participants to gain more information about the focus group method as well as 
for the clarification of some issues which may not have been addressed 
sufficiently in the discussion. Therefore, we suggest to take notes in 
this offline part.

\subsection{Post-processing}
In the post-processing phase the recorded material has to be
transcribed, analyzed, and evaluated.

The first analysis step is to check if the participants conformed to the
expected user types or whether the group has to be divided into
sub-groups (e.g., beginners and experienced users). Given this grouping, 
opinions expressed within the focus group can be associated with their user type 
during analysis, if applicable. 
This association allows to draw first conclusions for each user type. 

One method suitable for categorizing and extracting the information from 
the discussion is \emph{qualitative content analysis}~\cite{Mayring2002}.
Similarly to the classification of users, the material has to be 
categorized and opinions have to be assigned to the categories. The categories 
are based on the research question, the questions asked during the 
discussion, as well as the opinions given by the 
participants~\cite{MayringOverview}.

First, for each explicit and implicit question in the script, an own top-level 
category is defined, e.g., ``Strengths of the system related to the proof 
process''. Then the discussion is analyzed and for each opinion related to the 
top-level, if a suitable 
subcategory already exists, the opinion is assigned to that subcategory. If not 
then a new subcategory is introduced and the opinion is assigned to this 
category. For example, assume that the subcategory ``user interface'' was 
already defined and an opinion of one of the participants is: ``The user 
interface is great!''. Then this opinion would be assigned to the subcategory 
``user interface''. 

During this analysis, it is important to remain objective, to take all stated
opinions into consideration, and to avoid bias when
interpreting what has been said. It is useful to involve several
persons in this task, including the moderator.
When the material is categorized a revision of the categories may be done. For 
example some categories may be merged together to a larger or more 
abstract category.
After the categorization the opinions assigned to each of the subcategories 
have to be carefully analyzed and conclusions for the usability of this 
subcategory have to be drawn. This is a creative process and depends on the 
experience of the project members as well as the underlying tasks and research 
questions. Nevertheless, we give some suggestions, where to draw attention to 
during analysis.
It may be advisable to also take care which user 
type stated the opinion, as beginners often have different usability issues than 
intermediate or expert users. 
Also the reactions of the group should be taken into account, because an issue 
which the majority of the group agrees on might be an issue which the majority 
of the users in general might have as well. 
Attention should also be drawn to issues occurring with a higher frequency 
than others, regardless of the part or phase of the discussion they are 
expressed. There might be a correlation between the frequency and the relevance 
of an issue.

\section{The Script} \label{sec:script}

The main questions and tasks in the script were the same for both
discussions as we wanted to compare the results. The only differences
were adaptations of the questions and tasks to the different
terminology and adaptations of feature mock-ups to the specifics of
the two systems. The full scripts for our experiments are
available at \url{http://formal.iti.kit.edu/grebing/SWC} (as
the discussions were conducted in German, the original 
scripts are in German, but a translation to English is provided as well).
Table~\ref{fig:questions} gives a summary of the explicit questions for the 
participants. Our discussion was divided into three parts: the warm-up stage, the 
main stage and the cool-down stage. All three parts will be described in detail in 
the following. 

\begin{table}
\hspace*{\fill}\begin{tabular}{|p{0.02\textwidth}|p{0.98\textwidth}|}\hline
  \begin{turn}{-90}\textbf{Warm up}\end{turn}
  &
  \begin{enumerate}[leftmargin=*,partopsep=0pt,parsep=0pt,topsep=0pt, itemsep=0pt]
    \item Name typical use cases of the system.
    \item Name a strength of the system related to the proof process.
    \item Name a weakness of the system related to the proof process.
  \end{enumerate} 
  \\\hline
  \vspace{3.5cm}
  \begin{turn}{-90}\textbf{Main}\end{turn} &
  For the global and the local proof process:
  \begin{enumerate}[leftmargin=*]
    \itemsep0em
    \item How do you conceive a formalization/specification for a given problem?
      \begin{enumerate}
        \itemsep0em
        \item Please try to sketch the process. 
        \item Please point out steps of the process during which you get help/feedback from the system (if any).
        \item Do you repeat certain sequences of steps during the process? If so, please mark these loops.
      \end{enumerate}
    \item (Discussion)
      \begin{enumerate}
        \itemsep0em
        \item How do you rate the feedback you get from the system? (If 
negative: Where would be room for improvements?)
        \item Which steps of the process consume most of your time? Why?
        \item Which steps of the process annoy you? Could they be automated?
        \item What do you do if you get stuck?
        \item How do you rate the granularity of the proofs (in the local 
process)?
      \end{enumerate}
  \end{enumerate}

  For the mechanisms:
  \begin{enumerate}[leftmargin=*]
    \itemsep0em 
    \item Please describe the presented mechanism.
    \item Please rate the presented mechanism.
    \item What do you make of the approach?
  \end{enumerate}
  \\\hline
  \begin{turn}{-90}\textbf{Cool down}\end{turn} & 
  \begin{enumerate}[leftmargin=*]
    \item 
      Be creative and describe your ideal interactive proof system. Disregard 
      technical restrictions apart from the effectiveness of the automated proof 
      search. Name capabilities the system should definitely have. Name properties
      it must not have. 
  \end{enumerate}
  \\\hline
\end{tabular}\hspace*{\fill}

\caption{Summary of the questions from the script for the focus groups.}
\label{fig:questions}
\end{table}

\subsection{Warm-Up Tasks}
As warm-up task, we asked about typical application areas of the systems and 
about their strengths and weaknesses related to the proof process. Our intention 
for this part of the discussion was twofold. Firstly, we wanted the participants 
to slowly focus on the proof process of their system and ``warm up'' for the 
main part of the discussion. Secondly, our goal was to retrieve the advantages 
and disadvantages of the systems to draw conclusions about desirable features 
for interactive theorem provers. In addition, we expected to retrieve detailed 
information about the systems such that we can name the issues and give advices 
how to improve the systems.

\subsection{Main Part}
The main part of the discussion covered two topics: 
\begin{enumerate}
  \item 
    The proof process: What does the proof process look like? How does the tool 
    give support to the user during the process?
  \item 
    Mechanisms for understanding proof states: We confronted the participants 
    with mechanisms that might help them to understand the current state of 
    a proof during the proof process. 
\end{enumerate}

\xparagraph{Topic 1: The Proof Process.}
We divided the discussion for this topic into two parts, namely the
global proof process (finding the right formalization and decomposing
the proof task) and the local proof process (proving a single lemma or
theorem). For each part, participants were asked to describe their
typical proof process and discuss where the prover gives support and
where support is missing. We also asked what the most time-consuming
actions are.

By discussing the proof process, participants remember their typical 
interactions with the system in the past. This supports the subsequent 
discussion of how users get assistance from the systems during the 
proof process. Based on the participants' retrospection we hope to identify 
repetitive tasks or time consuming tasks, and parts where system feedback is missing. 

Also, we expect information on participants' use of the systems to solve particular tasks, 
on actions/phases or where they switch to other tools (e.g., text 
editors or pen and paper), on how they
inspect the proof state, and on how they guide the prover in finding a 
proof.
We also expect ideas from the participants on how and where they would improve the 
systems.

\xparagraph{Topic 2: Mechanisms for Understanding Proof States.}  For the second 
topic, we did not just ask for available or missing mechanisms. Instead we 
initiated a more focussed discussion by presenting mock-ups of mechanisms not yet 
built into the tools. 
The mock-ups were presented as a sequence of UI screenshots that have been 
modified according to the effect of a mechanism. These sequences of screenshots 
showed how to invoke the mechanisms and the corresponding effect of the 
mechanism\footnote{The screenshots are available at 
\url{http://formal.iti.kit.edu/~grebing/SWC/}.}.  
The purpose of the presented mechanisms was to support users
in understanding proof situations. The design of the particular
mechanisms was based on first hypotheses. First informal questionings of some 
users influenced the design as well.

The mock-ups included (a)~a mechanism for tracing formulas, terms, and variables 
that are generated during proof construction back to the original proof goal 
(for both tools), (b)~a visual support for proof management that shows which 
lemmas contribute to a proof (for Isabelle), and (c)~a mechanism for 
highlighting local changes between two adjacent nodes in the proof tree (for 
\KeY{}). Thus, we made use of the possibility to use focus groups to get a first 
assessment of new features.

For all presented mechanisms we had the same course of action and questions. 
First, the participants were asked to describe what they believe the mechanism 
does (i.e., the mechanism was not explained by the moderator). This was done 
both to avoid bias introduced by the moderator and to see if the mechanism is 
intuitive. Then, the participants were asked for their opinion on the usefulness 
of the mechanism. We expected to gain feedback about the presentation of the 
mechanism. If the participants had needed too much time to understand the 
functionality, we would have had to revise the presentation of the mechanism. 
We also had to some extent the opportunity to gain feedback for different 
presentations of the same functionality, such as the mechanism as own 
functionality with an own window or incorporated in the provers graphical user 
interface.  

With these mechanisms as starting point, we expected to get 
a discussion going about the usability problems w.r.t.\ the proof process where 
the presented mechanisms might help,
and to gain detailed insight into what
annoys the users and in which way they would like to see their system improve. 
Additionally, we expected answers to the question of which mechanism 
should be implemented first. 

\subsection{Cool-Down Task}
For the cool-down task, we asked the participants to be creative and
imagine their ideal interactive proof system.

The main idea behind the cool-down task is that the participants leave the 
discussion with a positive experience. Our intention was that we also gain some 
more, possibly creative, ideas on what features an ideal verification system should or 
should not incorporate.

\section{Results and Conclusion} 
\label{sec:results}

\subsection{Meta-level: Using Focus Groups to Evaluate ITPs}  
Our experiments clearly show that the focus-group method is not just
for business software but can be applied successfully to specialized
tools such as ITPs. We gained lots of insight from our experiments.

Focus groups are well suited for an explorative and qualitative
investigation of strengths and weaknesses in usability and the
usefulness of new features. They are particularly useful for systems
with a relatively small user base such as ITPs. Focus groups are a huge
step towards objective and reproducible experiments in the area of
usability, even though they do not provide precise quantitative data.
Another strength of focus groups is that participants voice ideas and discuss 
issues that they would not have talked about in single interviews, when 
topics are brought up by other participants. 
Our experience is that we gain detailed feedback to usability issues of the system. 
In addition, the discussions provide an understanding of how users use the 
system to achieve their goals.

The effectiveness of focus groups and their advantages over unstructured 
discussions, however, do not come for free. 
Conducting focus group discussions takes careful preparation
and is a non-trivial task. The effort and the 
required time is considerable. 
The approximate work-load of our experiments is given in
Table~\ref{table} (person-hours for both experiments, not
including the participants' and moderators' effort).
\begin{savenotes}

\begin{table}[tb]
  \caption{Time and  effort needed for conducting the two experiments 
(approximate).} \label{table}

\hspace*{\fill}%
\begin{tabular}{p{.33\textwidth}r@{\qquad}p{.33\textwidth}r} \hline
  script preparation &  20--40\phantom{.0} hours & 
    discussions &  5.5\phantom{.0} hours \\
  recruiting participants & 10\phantom{.0} hours & 
    transcription of recording &  20\phantom{.0} hours\\
  briefing moderators &  40\phantom{.0} hours  &
    analysis & 
\makebox[0pt][r]{80--160}\makebox[0pt][l]{}\phantom{.0} hours\\ 
  technical setup and testing &  8\phantom{.0} hours\\ \hline
\end{tabular}\hspace*{\fill}
\end{table}
 
\end{savenotes}

Our experiments support our hypothesis that the focus groups can also be 
performed successfully when computer
scientists are employed as moderators instead of hiring professional 
moderators. 
Their inexperience in moderating sometimes lead to biased questions and also
contributed to the discussions running over time. But their
familiarity with the basic terms of logic and theoretical computer
science allowed the discussions to proceed smoothly without spending
too much time on clarifying basic notions during the discussion.

We believe that it is possible to repeat the focus group discussions also 
for other theorem provers with the necessary preparation. The parties involved 
in preparing and conducting a focus group need some knowledge about the system 
under evaluation and should have some ideas where issues arise during the proof 
process. Also ideas for improvement are specific to each system and therefore
the involved parties should have worked out some ideas such that it is possible 
to provide mock-ups for the discussion. 
The experiments we have performed here can be repeated by using this 
description of our setup and the scripts available.

\subsection{The Kind of Insights to be Gained on the Concrete Level}
We identified strengths and weaknesses of the two systems, 
which mostly can be generalized to most or all ITPs. A typical weakness is, for example, 
an inadequate understanding of what the effect of automatic proof search 
strategies is. Users may lose the comprehension of the proof by applying 
automatic strategies, as in some cases the strategies do not give feedback 
which rules or transformations they apply and leave the user with a proof state 
that differs from the last seen state.

Also, technical issues that are annoying for the user and
compromise efficiency, were mentioned, e.g., unstable proof loading 
mechanisms or a user interface that is not sufficiently reactive. 
These answers point to where the systems' usability could be improved in particular. 

Discussions about
the proof process gave us insights into the feedback mechanisms of
both systems that support the proof process, e.g., the different automatic 
tools in Isabelle. Also issues that arise
during the processes have been mentioned by the participants and ideas
for improving certain aspects. These include for example presentations of the 
proof tree.  
We learned how the users use their
systems to accomplish certain proof tasks and where they switch to
other systems in order to get a better understanding of the current
proof state, for example by using an external text editor.  
By showing mock-ups of improvements we gained lively
feedback and opinions about the presented mechanisms. With these
opinions it should be possible for us to improve our mechanisms and
prototypically implement them. From the opinions we try to draw
conclusions about which mechanisms are more desired than others and
therefore get hints what to develop first.

The creative task at the end of the discussion lead to interesting and creative 
interaction mechanisms for ITPs, but also general desirable features for ITPs 
have been mentioned. Some of these features are already part of the systems;
others need improvement.

\section{Summary and Future Work}
\label{sec:conclusions}

To sum up, we have presented a method to qualitatively evaluate the usability 
of systems with a rather small user base. 
We have also shown, that it is possible to perform this evaluation without 
expert moderators, when being aware of this fact during the analysis of the 
transcribed material.
We made the experience that we gain insight onto the usability of ITPs using 
this method and that we can formalize first hypotheses about usability issues 
of the systems as well as a first lits of desirable features for ITPs.

A full analysis and interpretation of the recorded and transcribed material is
currently being done. This will result in a detailed report on
desirable features for interactive theorem provers. 

The mechanisms that attracted interest during the discussions need to be
further developed and prototypically implemented.
To ensure that the mechanisms suits the users needs and to evaluate 
whether they increase the usability, we will use
usability testing.
In addition, we will apply the User Experience Questionnaire method~\cite{ueq}
to assess the usability of the \KeY{}
system quantitatively. In this evaluation, we will determine, whether such
general-purpose questionnaires are helpful in evaluating the usability of
ITP systems, or whether more adaptable solutions are needed.

\subsection*{Acknowledgements}
We thank the participants in our focus group discussions on the
usability of \KeY{} and of Isabelle and, in particular, the two
moderators for their great work. In addition, we thank our project
partners from DATEV~eG for sharing their expertise in how to prepare
and analyze focus group discussions.
Florian B\"ohl was supported by MWK grant ``MoSeS''.

\bibliographystyle{eptcs}
\bibliography{refs_with_doi}

\end{document}